# Abundances, Ionization States, Temperatures, and FIP in Solar Energetic Particles

**Donald V. Reames**

IPST, University of Maryland, College Park, MD, USA

**Abstract** The relative abundances of chemical elements and isotopes have been our most effective tool in identifying and understanding the physical processes that control populations of energetic particles. The early surprise in solar energetic particles (SEPs) was 1000-fold enhancements in $^3$He/$^4$He from resonant wave-particle interactions in the small "impulsive" SEP events that emit electron beams that produce type III radio bursts. Further studies found enhancements in Fe/O, then extreme enhancements in element abundances that increase with mass-to-charge ratio $A/Q$, rising by a factor of 1000 from He to Au or Pb arising in magnetic reconnection regions on open field lines in solar jets. In contrast, in the largest SEP events, the "gradual" events, acceleration occurs at shock waves driven out from the Sun by fast, wide coronal mass ejections (CMEs). Averaging many events provides a measure of solar coronal abundances, but $A/Q$-dependent scattering during transport causes variations with time; thus if Fe scatters less than O, Fe/O is enhanced early and depleted later. To complicate matters, shock waves often reaccelerate impulsive suprathermal ions left over or trapped above active regions that have spawned many impulsive events. Direct measurements of ionization states $Q$ show coronal temperatures of 1 – 2 MK for most gradual events, but impulsive events often show stripping by matter traversal after acceleration. Direct measurements of $Q$ are difficult and often unavailable. Since both impulsive and gradual SEP events have abundance enhancements that vary as powers of $A/Q$, we can use abundances to deduce the probable $Q$-values and the source plasma temperatures during acceleration, ≈ 3 MK for impulsive SEPs. This new technique also allows multiple spacecraft to measure temperature variations across the face of a shock wave, measurements otherwise unavailable and provides a new understanding of abundance variations in the element He. Comparing coronal abundances from SEPs and from the slow solar wind as a function of the first ionization potential (FIP) of the elements, remaining differences are for the elements C, P, and S. The theory of the fractionation of ions by Alfvén waves shows that C, P, and S are suppressed because of wave resonances during chromospheric transport on closed magnetic loops but not on open magnetic fields that supply the solar wind. Shock waves can accelerate ions from closed coronal loops that easily escape as SEPs, while the solar wind must emerge on open fields.

## 1 Introduction

Relative abundances of the chemical elements and isotopes show striking variety in the populations of energetic particles we see, and they have given us the leverage to understand the physics of the particle origin. Examples include: (i) Dominance of energetic S and O in the Jovian magnetosphere points to sulfurous gasses emitted from volcanoes on the moon Io. (ii) Nearly pure H in the inner radiation belt of Earth results from the decay of neutrons produced in interactions of galactic cosmic rays (GCRs) with atoms in the Earth's atmosphere. (iii) The abundance of the rare elements Li, Be, and B in the GCRs themselves attest to the GCR interaction with interstellar H during their $\sim 10^7$ yr sojourn in the Galaxy. (iv) Anomalous cosmic rays (ACRs) contain only elements with high first ionization potential (FIP) that are neutral atoms in interstellar space so they easily penetrate the magnetic fields of the heliosphere to approach the Sun, where they are photo-ionized, picked up by the solar wind, and preferentially accelerated in the outer heliosphere, a complex journey. Where would we be without abundances? They give essential information and are signatures that identify populations of particles.

Probably the most unusual abundance pattern we know is in the little "impulsive" events of solar energetic particles (SEPs), the $^3$He-rich events. The abundance of the rare isotope $^3$He can be enhanced by a factor up to $10^4$ in these events. In a few events $^3$He exceeds H. Otherwise-rare heavy elements are also enhanced when compared with the corresponding abundances from the solar corona. First we found Fe/O enhanced a factor of ~10, and later we found the element abundances rising as a power law in their mass-to-charge ratio *A/Q* by a factor of ~1000 right across the periodic table from He up to Au and Pb. The impulsive SEP events have been associated with magnetic reconnection near the Sun in solar plasma jets that produce narrow, but relatively slow, coronal mass ejections (CMEs) and also accelerate the beams of electrons that produce type III radio bursts. It may be those electron beams that produce copious plasma waves at the gyrofrequency of $^3$He that are resonantly absorbed to selectively enhance energetic $^3$He. These events carry their characteristic abundance signature, like a beacon, wherever they appear. Meanwhile, we can use the strong power-law dependence on *A/Q*, with its associated pattern of charge states of the elements, to deduce an electron temperature of about 3 MK (1 MK $\equiv 10^6$ K) at the source of the ion acceleration. Abundances of impulsive SEP events are presented in Sect. 2.1.

By way of contrast, the average abundances of elements in the large "gradual" SEP events, like those in the solar wind, provide a measure of element abundances in the solar corona. Coronal abundances differ from those in the solar photosphere as a bilevel function of the first ionization potential (FIP) of the ion. Elements with low FIP (<10 eV) such as Mg, Si, and Fe, are ionized in the photosphere and chromosphere and are more easily swept up into the corona than high-FIP elements, like He, O, and Ne, that begin as neutral atoms. Thus the low-FIP elements are a factor of ~3 more abundant. All atoms become highly ionized at the ~1 MK temperature of the corona where they are available to be accelerated by shock waves driven out from the Sun by fast, wide CMEs to produce gradual SEP events. Scattering of SEPs streaming out from the source depends upon particle rigidity, or upon *A/Q* at a given velocity, so that Fe scatters less than O, for example. This creates regions of enhanced Fe/O where Fe has forged ahead and regions of depleted Fe/O behind.  This time dependence is converted into longitude dependence by solar rotation.  While these regions may average to give the coronal abundance, the individual regions provide a pattern of $Q$-dependence that again allows us to determine a source plasma temperature, usually ~1 – 2 MK where about half the electrons have been removed from Fe.   In some active regions, shock waves also reaccelerate suprathermal ions left over from previous *impulsive* SEP events.  Then a component of the $^3$He-rich, Fe-rich, 3 MK plasma shines through.  Abundances in gradual events are discussed in Sect. 2.2

Measurements of ionization states of heavier ions such as Fe can tell us the typical plasma temperature from which they came and thus help identify their origin.  Such measurements have been made directly in instruments, and by using the geomagnetic field.  In either case, deflection of particles of a given velocity by electric or magnetic fields in an instrument depends upon their mass-to-charge ratio $A/Q$.  Unfortunately, the instruments are complex, are limited to low energy, and are rarely available, while the geomagnetic measurements are local and are not possible throughout the heliosphere.

However, the physical processes of SEP acceleration and transport *themselves* involve electric and magnetic fields, so the relative enhancement and suppression of different species at a given velocity *also* depends upon $A/Q$.  Conveniently for us, element abundances are frequently altered as powers of $A/Q$.  Why not use the $A/Q$-dependent element abundances nature has already provided to determine $Q$-values and electron temperatures of the source plasma?  Perhaps the electric and magnetic fields near the particle's source can analyze the ions for us just as well as the field of the Earth or those in instruments. As a bonus, the ionization states deduced are often those at the original source, unaltered by later stripping during transit.  Unfortunately the new technique does not measure distributions of $Q$ and temperature, so direct measurements are still important, but having extended coverage of mean values of $Q$ and $T$ for many more SEP events is a great benefit.  Ionization states, especially these new techniques, are discussed in Sect. 3 and variations in the source abundances of SEPs, especially He, in Sect. 4.

Thus we review details of measurements of SEP abundance and charge states and the physical processes those measurements have helped us understand.  It is no surprise that abundances have already contributed significantly to our understanding of SEPs. This gives us a basis to describe the new method of using power-law abundance enhancements to determine source plasma temperatures and the early findings from that technique.  Having resolved questions about He, we then turn to the "FIP-effect" and the SEP view of the corona, as compared with that of the solar wind.  This comparison shows differences in the abundances of C, P, and S in SEPs and in the solar wind in Sect. 5. These differences can be understood by the theory of ponderomotive forces of Alfvén waves on the transport of ions through the chromosphere on open (solar wind) and closed (SEPs) magnetic field lines. Thus there are differences in the regions where SEPs and the solar wind can be selected.  Shock waves can sample ions from closed, but weak, high coronal loops to generate SEPs that easily escape, but the solar wind must come only from field lines were open from the chromosphere upward.

In Sect. 6 we return to ionization states and temperatures and discuss early results in mapping them in space and time across an accelerating shock wave, followed by a general discussion of relevant consequences (Sect. 7) and summary (Sect. 8).

## 2 SEP Abundances

### *2.1 Impulsive SEP events*

$^3$He-rich events were first observed by Hsieh and Simpson (1970). To a community familiar with nuclear fragmentation of GCRs they were first thought to be produced by nuclear reactions in solar flares, but there was no evidence of $^2$H, $^3$H, or other reaction secondaries. Then Serlemitsos and Balasubrahmanyan (1975) found $^3$He/$^4$He = 1.52 ± 0.10, compared with (4.08 ± 0.25) × 10$^{-4}$ in the solar wind (Gloeckler and Geiss 1998), but they also found $^3$He/$^2$H > 300. How could so much $^3$He be produced in nuclear reactions with no $^2$H or $^3$H? Yet, there is certainly evidence of nuclear reactions in the corona in flares from the neutrons (Evenson et al. 1983, 1990) and γ-ray lines (e.g. Ramaty and Murphy 1987; Kozlovsky, Murphy, and Ramaty 2002) we see from the footpoints of magnetic loops during flares. However, isotopes of Li, Be, and B, also expected from the γ-ray lines, have never been observed in SEPs in space. Limits on Be/O or B/O in large SEP events are < 2 × 10$^{-4}$ (e.g. McGuire, von Rosenvinge, and McDonald 1979; Cook, Stone, and Vogt 1984). Neutrals from nuclear reactions escape, but charged particles cannot escape the closed magnetic loops of solar flares.

Thus the ions we see in space are accelerated in reconnection events on the open magnetic fields of solar jets (e.g. Kahler, Reames, and Sheeley 2001; Reames 2002) rather than those on the closed loops of flares. Furthermore, the huge enhancement of $^3$He is not produced in nuclear reactions at all, but by preferential acceleration in resonant wave-particle interactions at the gyrofrequency of $^3$He. A variety of preferential heating mechanisms were suggested (*e.g.* Fisk 1978) but acceleration remained elusive. A spectrum of waves would be absorbed at the gyrofrequencies of the most abundant species in the plasma, $^1$H and $^4$He, but the much rarer $^3$He could absorb considerable energy at its gyrofrequency, which lies alone between those of the two most abundant species, without significantly damping the waves there.

Early observations found non-relativistic electron events in space (*e.g.* Lin 1974) that were associated with small impulsive X-ray events at the Sun and with the production of metric type III radio bursts. Lin thought many of these to be "pure" electron events, lacking observable protons or other ions. However, $^3$He-rich events were subsequently associated with these electron events (Reames, von Rosenvinge, and Lin 1985) and directly with both metric and kilometric type III radio bursts (Reames and Stone 1986). Later, it was suggested that these streaming electrons produced electromagnetic ion-cyclotron waves that resonated with mirroring $^3$He ions to produce the striking enhancements in $^3$He-rich SEP events (Temerin and Roth 1992; Roth and Temerin 1997). More recently, Liu, Petrosian, and Mason (2006) have been able to fit the complex spectra of $^3$He and $^4$He (Mason 2007).

With improving measurements, it was found that heavy-ion abundances, such as Fe/O, were also enhanced in $^3$He-rich events (*e.g.* Mason et al. 1986; Reames, Meyer, and von Rosenvinge 1994) and Fe/O became a more reliable indicator of impulsive events than $^3$He/$^4$He, since the latter varies strongly with energy while Fe/O is better behaved (Mason 2007). In fact, with modern data, when we simply look at Ne/O vs. Fe/O in all 8-hr periods during 19 years, we find the distribution in Figure 1. Data in this figure are unbiased with respect to time or event selection, although, clearly, individual gradual events occupy many more time periods, and many small impulsive events with too few ions are overlooked. This bimodal (Reames 1988) abundance distribution helps resolve the populations of impulsive and gradual SEP events.

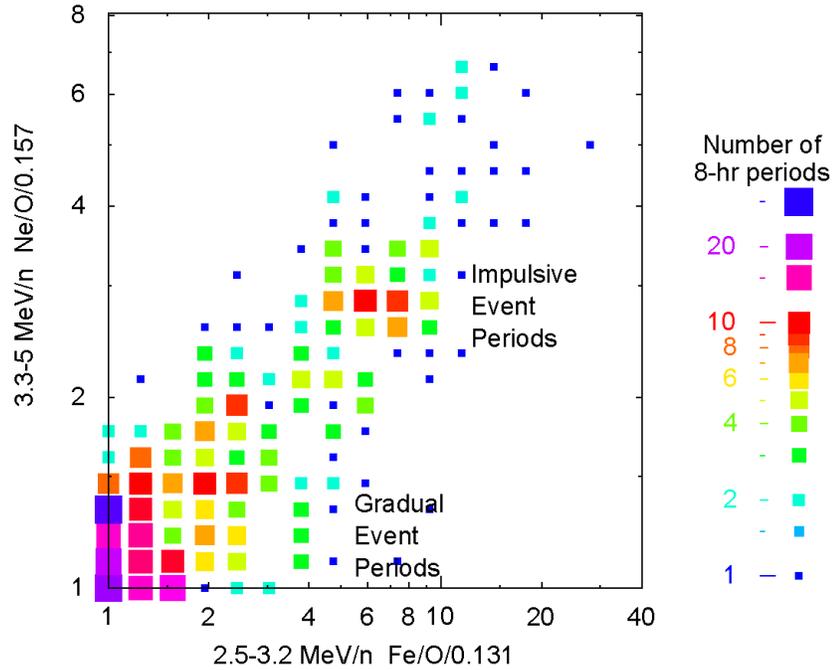

**Fig. 1** Values of enhancements of Ne/O *vs.* Fe/O near 3 MeV amu$^{-1}$ are binned for all 8-hr intervals during 19 years which have errors of 20% or less. The cluster of periods near the origin at (1, 1) represents gradual SEP event periods, that determine the normalization factors. The peak near (7, 3) is from the impulsive SEP events.

Eventually observations were extended to the rest of the periodic table above Fe (Reames 2000; Mason et al. 2004; Reames and Ng 2004), although *not* with single-element resolution. The enhancement of these heavy ions, relative to their coronal abundance rises as a power of $A/Q$, with $Q$-values defined at a temperature of ~3 MK, rising three orders of magnitude between He and Pb (Reames 2000; Mason et al. 2004; Reames and Ng 2004; Reames, Cliver, and Kahler 2014a) as seen in Figure 2.

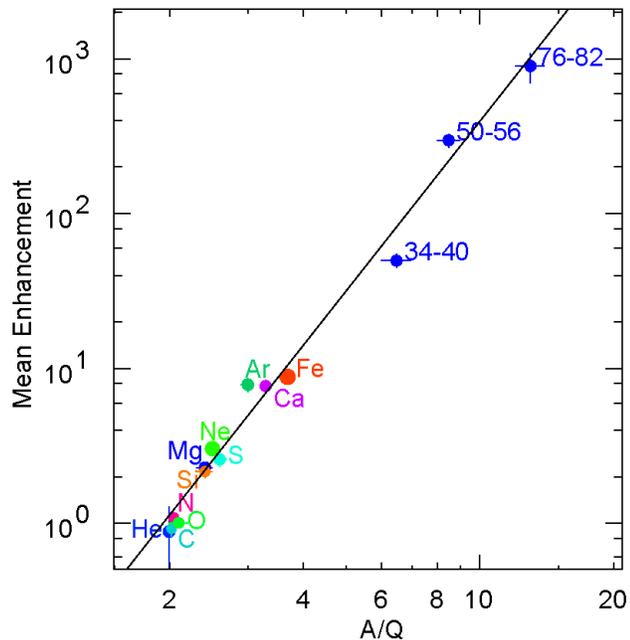

**Fig. 2** The mean enhancements in the abundances of elements in impulsive SEP events, relative to reference "coronal" abundances defined in gradual SEP events, is shown as a function of $A/Q$ of the element with $Q$–values defined at ~3 MK. Elements are identified by name or by intervals of atomic number $Z$. For the least-squares fit line shown in the figure the enhancement varies as the 3.64 ± 0.15 power of $A/Q$ (Reames, Cliver, and Kahler 2014a).

The recent measurements (Reames, Cliver, and Kahler 2014a) strongly associate the impulsive events with slow, narrow CMEs supporting their origin in solar jets, originally suggested by Kahler, Reames, and Sheeley (2001, see also Reames 2002 and review of jets by Raouafi et al. 2016). The CMEs associated with impulsive SEP events rarely drive shock waves since they move along *B* at less than the Alfvén speed (Vrsnak and Cliver 2008), thus they lack radio type II bursts, and their mean width is 75$^o$ *vs.* >130$^o$ for those associated with gradual SEP events. The association with jets is strengthened by the presence of beamed electrons and type III-burst association of impulsive SEP events. The enhancement of the heavy elements as a function of *A/Q* appears in particle-in-cell simulations (e.g. Drake et al. 2009) to result directly during the collapse of islands of magnetic reconnection also expected to occur in the solar jets. Drake et al. (2009) find that the rate of production of ions *dN/dt* is

$$\frac{dN}{dt} \propto w^{3-\alpha} \propto \left(\frac{A}{Q}\right)^{\alpha-3} \quad (1)$$

where the distribution of islands of width *w* is $w^{-\alpha}$. Other properties of impulsive SEP events have been reviewed by Mason (2007) and by Reames (2017a).

## *2.2 Gradual SEP events*

The first SEP events observed were sufficiently large for GeV protons to produce fragments from nuclear interactions in the Earth's atmosphere that were detectible at ground level (Forbush 1946). These ground-level enhancements (GLEs) revealed nothing whatever about the abundances of the elements in the impinging SEPs. Nevertheless, even in the earliest sounding rocket observations of C, N, and O (Fichtel and Guss 1961) and of Fe (Bertsch, Fichel, and Reames 1969) and later measurements in space (e.g. Teegarden, von Rosenvinge, and McDonald 1973), authors attempted comparisons with abundances of the solar corona. However, with increasing measurements of SEP abundances in space, it was found that the abundances, averaged over many SEP events, divided by the corresponding solar photospheric abundances (knowledge of which was also evolving) showed a difference of a factor of ~3 in the levels of elements with low and high FIP (Webber 1975; Meyer 1985). $^3$He-rich events were explicitly excluded from this average by Meyer. The individual SEP events forming the average showed a "mass-bias" or a dependence on *A/Q* according to Meyer (1985).

With improving instruments, average abundances from gradual SEP events (Reames 1995, 2014) are a standard for comparison with the corona and solar wind (Mewaldt et al. 2002; Schmelz et al. 2012). Figure 3 shows a recent FIP plot of average element abundances in SEPs (Reames 2014, 2017a, 2018) divided by the corresponding photospheric abundances where the dominant element species are from Caffau et al. (2011) supplemented by other elements from Lodders, Palme, and Gail (2009). The difference in FIP levels arises because the high-FIP elements are neutral atoms in the photosphere while the low-FIP elements are ions. The ions are more easily conveyed into the corona by the action of Alfvén waves, for example (e.g. Laming 2009, 2015)

A comparison of the FIP dependence using the alternative photospheric abundances of Asplund et al. (2009) is shown by Reames (2015). SEP He abundances from events with different source plasma temperatures are shown in Figure 3 and are discussed in Sect. 4.1. We will compare FIP plots for SEPs with that for the solar wind in Sect. 5. Whenever we talk about "enhancements" in this paper, we mean observed abundance ratios divided by these average SEP abundances, i.e. the SEP corona.

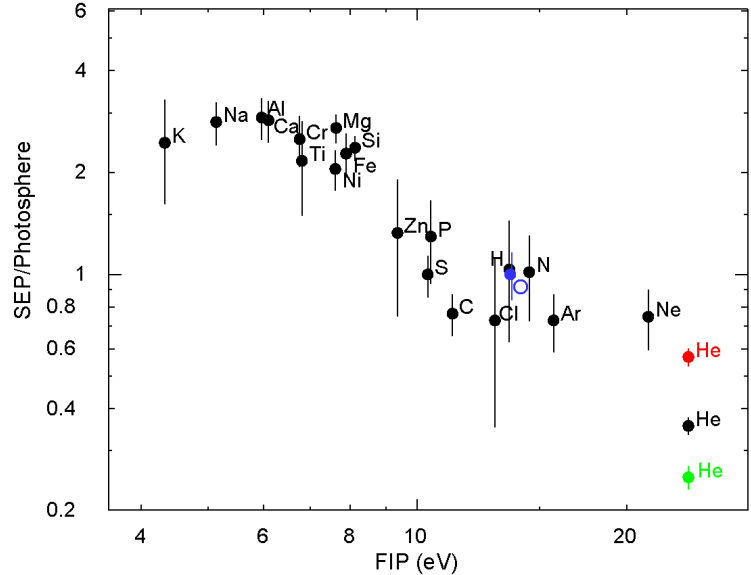

**Fig. 3** Element abundances from SEPs (Reames 2014, 2017a) are divided by photospheric abundances (Caffau et al. 2011, Lodders, Palme, and Gail 2009), normalized at O (blue). Abundances of He in three event regions correspond to He/O= 90 (red), 57 (black) and 40 (green) (see Sect. 4.1).

Direct measurement of the charge state $Q$ for a variety of ions is difficult and is generally limited to energies below 1 MeV amu$^{-1}$. The earliest measurement of $Q$-values for SEP ions (Luhn et al. 1984) were used with element abundances by Breneman and Stone (1985) to show SEP events with enhancements that either increased or decreased as a power-law function of $Q/A$. These events are shown in Figure 4. Values of $Q$ of Fe from the measurements of Luhn (1984) showed a distribution peaking near $Q_{Fe} \approx 14$ which corresponds to $T \approx 2$ MK that controls the groupings of the elements along the top of Figure 4. More-recent fits to the $A/Q$ dependence of abundance enhancements during early and late 8-hr periods in the SEP event of 8 November 2000 are shown in the right panel of Figure 4 (Reames 2016a).

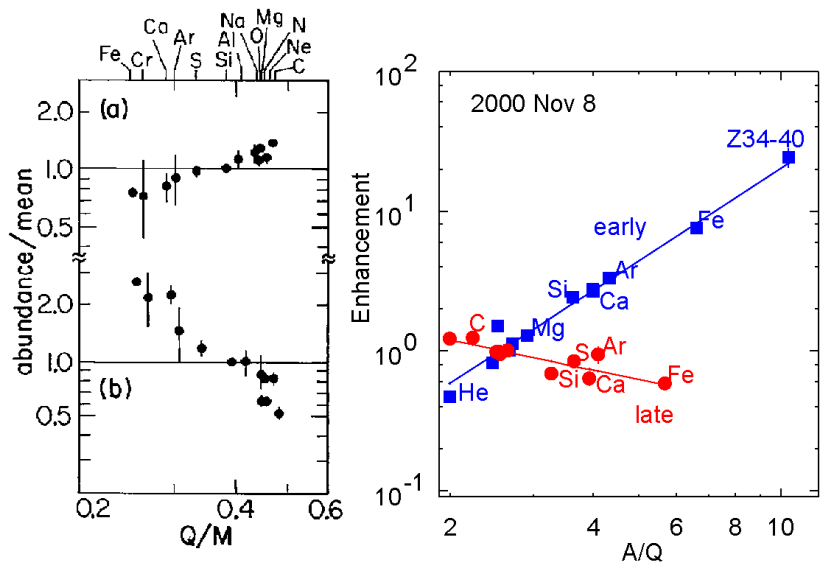

**Fig. 4** The left panel shows the dependence of elemental abundances on the charge-to-mass ratio $Q/M$ (our $Q/A$) of the elements is shown for two large SEP events by Breneman and Stone (1985). The right panel shows the best-fit power-law dependence on $A/Q$ during early and late time periods in the 8 November 2000 SEP event (Reames 2016a).

The power-law dependence on *A/Q* found by Breneman and Stone (1985) actually came as a surprise to the SEP community. However, this dependence comes directly out of the SEP transport theory introduced by Parker (1963). If we assume that the scattering mean free path $\lambda_X$ of species X depends upon as a power law on the particle magnetic rigidity *P* as $P^\alpha$ and upon distance from the Sun *R* as $R^\beta$ we can use the expression for the solution to the diffusion equation (Equation C3 in Ng, Reames, and Tylka 2003 based upon Parker 1963) to write the enhancement of element X relative to O as a function of time *t* as

$$X/O = L^{-3/(2-\beta)} \exp\{(1-1/L)\tau/t\} r^S \qquad (2)$$

where $L = \lambda_X/\lambda_O = r^\alpha = ((A_x/Q_x)/(A_O/Q_O))^\alpha$ and $\tau = 3R^2/[\lambda_O(2-\beta)^2 v]$ for particles of speed *v*. Since we compare ions at a fixed velocity, their rigidities are replaced by the corresponding values of *A/Q*. The factor $r^S$ represents any *A/Q*-dependent power-law enhancement at the source, prior to transport. For shock acceleration of impulsive suprathermal ions, it describes the power-law enhancement of the seed particles (i.e. Equation 1). For shock acceleration of the ambient coronal material, $S = 0$.

We achieve a power-law approximation if we expand $\log x = (1-1/x) + (1-1/x)^2/2 + \ldots$ (for $x > \frac{1}{2}$). Using only the first term to replace $1-1/L$ with $\log L$ in Equation (2), we have

$$X/O \approx L^{\tau/t - 3/(2-\beta)} r^S \qquad (3)$$

for $L > \frac{1}{2}$, as an expression for the power-law dependence of enhancements on *A/Q* for species X.

More generally, we can write Equation (2) in the form $X/O = r^p$, where the power *p* is linear in the variable $1/t$, so that

$$p = \alpha\tau/t + S - 3\alpha/(2-\beta). \qquad (4)$$

This time dependence is directly measurable from the SEP-abundance observations (Reames 2016b).

Upsetting the simple picture of impulsive and gradual SEP events is the fact that fast shock waves can also reaccelerate residual suprathermal ions left over from previous impulsive SEP events. In fact these suprathermal ions may even be preferentially accelerated. Mason, Mazur, and Dwyer (1999) first observed small, but significant, enhancements in $^3$He in SEP events that were otherwise large gradual events, and strong enhancements in $^3$He/$^4$He are now seen even in quiet periods (Desai et al. 2003; Bučík et al. 2014, 2015; Chen et al. 2015). These $^3$He-rich periods must involve many small jets in active regions producing many impulsive SEP events that are too small to be resolved individually. Shock waves passing through these regions may preferentially accelerate the suprathermal ions, depending upon the angle between the field *B* and the shock normal (Tylka et al. 2005; Tylka and Lee 2006). Quasi-perpendicular shocks may preferentially accelerate the faster suprathermal ions that can overtake the shock more easily from downstream. A complex variation of abundance ratios, such as Fe/C, occurs especially at high energies > 10 MeV amu$^{-1}$ where spectral breaks depend upon both the shock geometry and *A/Q*, which differs between the ambient and the suprathermal ions (Tylka et al. 2005; Tylka and Lee 2006). These reaccelerated impulsive suprathermal ions can have the spectra and intensities of shock acceleration, seen in other gradual SEP events, but the somewhat-diluted abundances of impulsive SEP events.

The association of gradual SEP events with CME-driven shock waves began with the 96% correlation between large SEP events and fast, wide CMEs (Kahler et al. 1984), and the broad spatial distribution of abundances (Mason, Gloeckler, and Hovestadt 1984), and continued with shock theory (Lee 1983, 2005; Zank, Rice, and Wu 2000; Zank, Li, Verkhoglyadova 2007; Ng and Reames 2008), reviews of properties (Gosling 1993, Reames 1999, 2013, 2015, 2017a; Kahler, 1992, 1994, 2001; Cliver, Kahler, and Reames, 2004; Cliver and Ling, 2007; Gopalswamy et al., 2012; Mewaldt et al. 2012; Lee, Mewaldt, and Giacalone, 2012; Cliver 2016; Desai and Giacalone 2016) and the detailed spatial CME studies of Rouillard et al. (2011, 2012, 2016).

## *2.3 What about flares?*

Solar flares exist precisely because energetic particles in them are magnetically trapped. Without containment there would be minimal flash and X-rays; energetic charged particles would mostly be magnetically mirrored away from the Sun with minimal flash, as seen in jets where a small trapping region does exist (Shimojo and Shibata 2000; Reames 2002, 2017a). Particles accelerated during magnetic reconnection on closed loops in flares scatter into the loss cone and plunge into the low corona where they interact in the denser matter and come to rest. Electron bremstrahlung produces hard X-rays, and energetic ions undergo nuclear reactions producing $\gamma$-rays and neutrons that escape into space, but the ions do not. As energetic particles lose energy and stop in the dense plasma, their energy deposit produces heating up to 10 – 20 MK and the hot plasma evaporates back up into the loops, causing the flash. Analysis of $\gamma$-ray lines from large flares suggests that the accelerated "beam" is $^3$He-rich (Mandzhavidze et al. 1999; Murphy, Kozlovsky, and Share 2016) and Fe-rich (Murphy et al. 1991), just like the impulsive SEP events that escape into space from jets. However, although $^2$H, $^3$H, Li, Be, and B secondary ions are produced in flares, just as they are in the GCRs, these charged secondary ions cannot escape the flares. Nor can the primary ions. Even during Carrington's (1860) first observation of a flare, he was "surprised … at finding myself unable to recognize any change whatever as having taken place." The fields maintain their shape. Modern instruments allow measurement of the reconnection magnetic flux and a recent data base contains reconnection flux for 3137 solar-flare ribbon events (Kazachenko et al. 2017). However, large reconnection events lacking fast shock waves produce beautiful flares, but *no* SEPs (Kahler et al. 2017). Flares require containment by closed magnetic fields and disrupt those fields very little; the energy of particles accelerated in flares is dissipated by heating in the footpoints and loops and the particles are not released into space.

In principle, $^3$He-rich, Fe-rich, gradual events might have been seeded by flares, but the SEPs are too cool, 3 MK, and they have no nuclear-reaction fragments. As stated previously, none of the secondary ions expected from reactions are seen in space. Despite the enhancements of $^3$He, the isotopes $^2$H and $^3$H are not observed in SEPs. Limits on Be/O or B/O in large SEP events are $<2 \times 10^{-4}$ and Li/O < 0.001 (e.g. McGuire, von Rosenvinge, and McDonald 1979; Cook, Stone, and Vogt 1984). Furthermore, we do not see hot plasma. At 20 MK, Ne, Mg, and Si would be fully ionized and have $A/Q$ = 2, like O. In gradual SEP events we do see plasma at ~3 MK that is consistent with reacceleration of impulsive suprathermal ions from jets. These events are characterized by enhancements of Ne/O, unlikely for fully ionized Ne and O.

Furthermore, the classical magnetic topology of eruptive events suggests that major flaring produced by reconnection associated with the ejection of a CME would occur *beneath* or *behind* the CME in eruptive flares. If we wanted to look for residue from a flare among SEPs, we should look late in an event, *after* passage of the CME, for Be/O and B/O, for unenhanced Ne/O and Mg/O, and for a source plasma temperature above 10 MK. Any small leakage of SEPs from flares is apparently dwarfed by SEPs accelerated at the CME-driven shock that have filled the reservoir (e.g. Reames 2013, 2017a) behind the shock.

Without containment there would be no flare. That containment prevents us from seeing SEPs accelerated at reconnection sites in flares. However, we do often see the ions from reconnection events that produce jets on open field lines nearby. There certainly may be events involving reconnection on *both* open and closed field lines. In those events acceleration on closed field lines produces a flare, while acceleration on open field lines produces a jet and the SEPs that are visible in space, but abundances show that the open and closed regions are not connected and do not communicate. Logically, it is not possible for the reconnection of closed field lines with other closed field lines to produce open field lines. There is simply no evidence of a flare contribution to SEPs.

## *2.4 SEP Resolution*

Especially for the determination of coronal element abundances, it is actually a virtue that most modern SEP telescopes are totally insensitive to ionization states and that they treat signals produced by elements C through Fe identically. In contrast, a spectral line measurement determines a single ionization state of a single element and typical solar wind instruments must resolve each ionic state separately and sum them to produce abundances suitable for a FIP plot like that in Figure 3.

A study of SEP telescopes is beyond the scope of this review (see Chapt. 7 of Reames 2017a), but modern Si "solid-state" telescopes are quite simple. Logically, they consist of a thin detector, a thick detector and an anti-coincidence detector. Particles that traverse the thin detector and stop in the thick one will deposit energy in accordance with the well-known range-energy relation in Si. The telescopes are calibrated before flight at accelerators with beam of not only O and Fe, but also Ag and Au in some cases. The telescopes can "saturate" at high intensities when more than one particle can enter the telescope within its response time, typically a few microseconds, but otherwise the instruments typically provide reliable, stable, and unbiased readout at their design resolution for over 20 years or more. Elements fall on well-defined response "tracks" and the presence of any background can be seen between them. Instrument resolution is shown for the observations discussed in this review on the spacecraft ISEE 3 (Reames 1985), Voyager (Breneman and Stone 1985), Wind (von Rosenvinge et al. 1978; Reames 2000, 2014), ACE (Stone et al. 1998; Mason et al. 1998; Leske et al. 2007)), STEREO (Mewaldt et al. 2008).

The price we pay for highly reproducible abundance measurements is that we must disentangle FIP dependence from power laws in *A/Q*, and possibly other variations. This FIP dependence is determined at the base of the corona and the *A/Q* dependent power laws during acceleration (impulsive events) or transport (gradual events). If we want to use abundances to measure source plasma temperatures, we must use the power law dependence of *A/Q* to determine *Q* and hence temperature.

# 3 Ionization States and Temperatures

## 3.1 Direct and Geomagnetic Measurement

The earliest direct measurements of ionization states (Luhn et al. 1984, 1987) provided a clear difference between gradual and impulsive events. On average, gradual events had $Q_{Fe}$ =14.1 ± 0.2 and $^3$He-rich events had $Q_{Fe}$ =20.5 ± 1.2 with wide distributions in individual events. The value for impulsive events was sometimes interpreted as originating in hot solar flares. However, it was later found that the value varied with ion energy as expected from the equilibrium obtained by passing the ions through a small amount of matter (DiFabio et al. 2008), consistent with sources below about 1.5 $R_S$. The matter traversed is not adequate to cause significant energy loss for the heavy elements, since their power law in *A/Q* (Figure 2) has not been disrupted, even at very high *Z*.

The higher intensities and energies in gradual events permitted using deflection in the geomagnetic field to measure charge states (Leske et al. 1995) at 15 – 70 MeV amu$^{-1}$ where the average $Q_{Fe}$ = 15.2 ± 0.7. Meanwhile, Tylka et al. (1995) found $Q_{Fe}$ = 14.1 ± 1.4 at 200 – 600 MeV amu$^{-1}$. More recently, Klecker (2013) summarized mean values of $Q_{Fe}$ for gradual events as being in the range from 10–14. These values are consistent with coronal temperatures of ~1.2–2.5 MK, varying from event to event.

## 3.2 Inferring Temperatures from Abundances

The use of abundances alone to imply temperatures began when Reames, Meyer, and von Rosenvinge (1994) noted that He, C, N, and O in impulsive SEP events all had coronal abundances, as in the average of gradual events, thus they were probably all fully ionized and had the same value of *Q/A*=0.5. Meanwhile, Ne, Mg, and Si all had very similar enhancements while the enhancement of Fe was larger. O is fully ionized at a temperature of *T* > 3 MK, while Ne, Mg, and Si have closed shells of two orbital electrons with similar values of *Q/A* ~ 0.43 for 2.5 < *T* < 4 MK. In this temperature range, 3 < *T* < 5, Fe has *Q/A* ~ 0.28. Using tables of *Q vs. T*, mainly from Arnaud and Rothenflug (1985), the pattern of abundances suggested this temperature region.

With more accurate measurements it became clear that the enhancements of Ne/O > Mg/O >Si/O, opposite the expected ordering in *Z*, which occurs for the respective values of *A/Q* at *T* = 2.5 – 3.2 MK (Reames, Cliver, and Kahler 2014a). This together with the availability of enhancements for elements 34 ≤ Z ≤ 82 led to the plot of enhancement *vs. A/Q* in Figure 2.

The theoretical values of *A/Q vs. T* for elements that we use to derive temperatures are shown in Figure 5. Values of average *Q vs. T* for elements up to Fe are from Mazzotta et al (1998); those for typical elements above Fe come from Post et al (1977). The pink "active region" band in the figure marks the region where the observed pattern of element enhancements in impulsive SEP events matches the pattern of *A/Q*; the *A/Q* values used in Figure 2 come from this band in Figure 5.

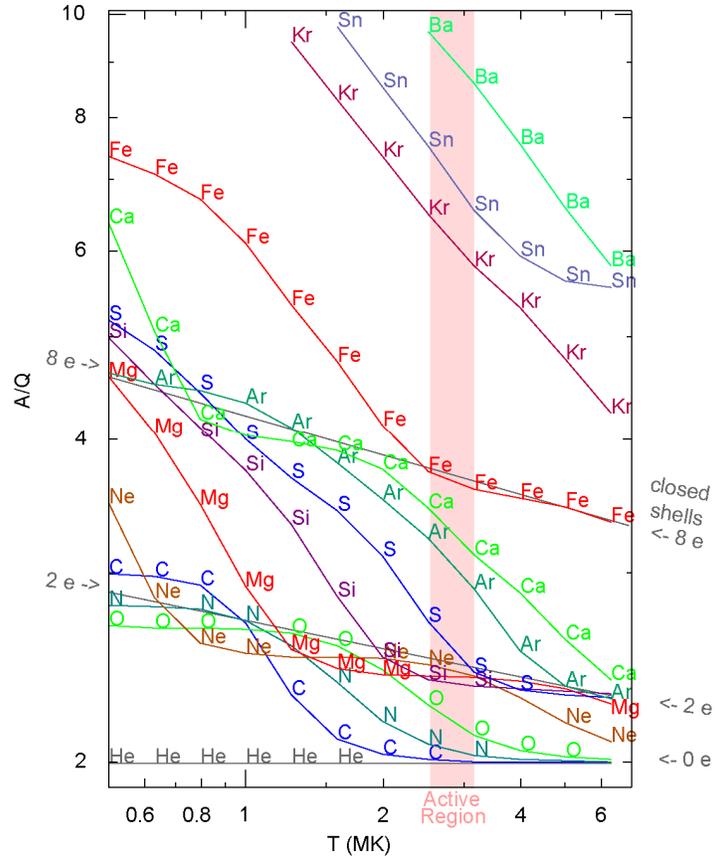

**Fig. 5** *A/Q* is shown as a function of the theoretical equilibrium temperature for elements named along each curve. Data are from Mazzotta *et al.* (1998) up to Fe and from Post *et al.* (1977) above. Points are spaced every 0.1 unit of $\log_{10} T$. *A/Q*-values tend to cluster in bands produced by closed electron shells; those with 0, 2, and 8 electrons are shown, He having no electrons. Elements move systematically from one band to another as temperatures change. The shaded pink region corresponds to active-region temperatures found for impulsive SEP events.

### 3.2.1 Impulsive SEP events

If we wish to determine a temperature for an individual SEP event, we simply fit the observed abundance enhancements *vs.* the *A/Q*-values at *each* temperature that could be of interest and record the value of $\chi^2/m$ *vs. T*, where *m* is the number of degrees of freedom of the fit, i.e. the number of elements measured minus two. The best fit and corresponding temperature for each time are chosen at the minimum value of $\chi^2/m$.

Figure 6 shows data and power-law least-squares fit lines using *A/Q*-values at three temperatures for the impulsive SEP event of 1 May 2000. This event has been studied extensively in other respects (e.g. Reames, Ng, and Berdichevsky 2001; Kahler, Reames, and Sheeley 2001; Reames, Cliver, and Kahler 2014a). The best fit at 3.2 MK (red circles), seen in the lower panel of Figure 6, is somewhat distorted by the high value of Ne/O. High Ne/O is common in impulsive SEP events, but these events may also have high non-thermal variations of ~30% in other species, as well (Reames, Cliver, and Kahler 2015; Reames 2016b). Decreasing the temperature from 3.2 to 1.6 MK (blue filled circles), increases *A/Q* for N, and O, but not for He, and C, and increases *A/Q* for Si through Fe but not for Ne and Mg; this causes severe distortion. Increasing the temperature from 3.2 to 6.3 MK (green squares), moves Ne down toward O, despite its greater enhancement and reduces *A/Q* greatly for Ca, but not for Fe, again causing obvious distortion. These changes in *A/Q* can be followed in Figure 5. Intermediate values in *T* result in intermediate values of $\chi^2/m$.

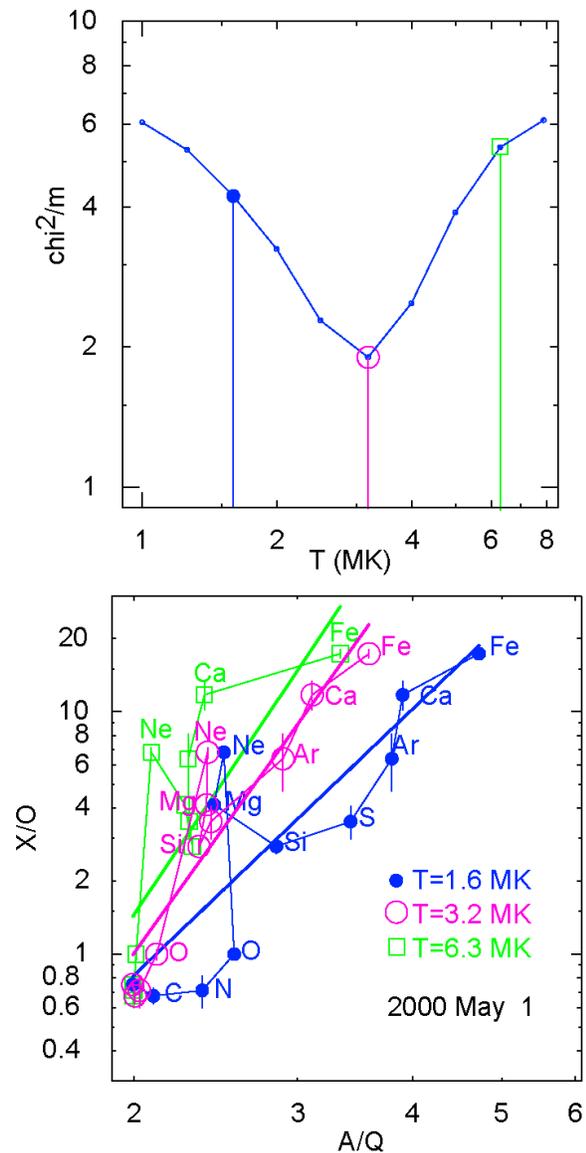

**Fig. 6** Least-squares power-law fits are shown for the impulsive SEP event of 1 May 2000. The lower panel shows data and fits of element abundance enhancements, relative to coronal values, X/O as a function of $A/Q(T)$ for three selected temperatures. The observed values of X/O do not change with temperature. The upper panel shows $\chi^2/m$ vs. $T$ for fits at ten temperatures with values for the three selected fits highlighted with the corresponding symbol. The value of $\chi^2/m$ for the best fit at $T \sim 3.2$ MK remains above 1 largely because of the excess enhancement of Ne, but the fits become much worse and the data patterns are distorted as the assumed temperatures are raised or lowered.

　　　　The original study of 111 individual impulsive SEP events (Reames, Cliver, and Kahler 2014b) found 108 events had source plasma temperature minima of either 2.5 or 3.2 MK. A subsequent study (Reames, Cliver, and Kahler 2015) found some variation when weighting for the fits allowed for non-thermal variations up to 30% in quadrature with the statistical error, but a search found a dearth of events with abundance patterns that would be expected outside the 2 – 4 MK range. Individual impulsive SEP events do not stray far from the average.

### 3.2.2 Gradual SEP Events

The observations of Breneman and Stone (1985), shown in Figure 4, provided the first strong evidence of a power-law dependence of abundance enhancements on $A/Q$, but the possible effect of temperature dependence of $Q$ on the pattern of elements at the top of the plot in Figure 4 was not fully appreciated. In subsequent years, direct and geomagnetic measurements of $Q$ became available that began to show the variability of $Q_{Fe}$ and the temperatures that it implied. However, these measurements soon became less common and systematic studies of large samples of events were rare.

The success of the measurements of a large sample of impulsive SEP events using abundance enhancements *vs.* *A/Q* suggested the use of the same technique for gradual SEP events and it was shown that a power-law dependence of abundance enhancements upon *A/Q* was expected from diffusion theory (Reames 2016a, 2016b). Gradual events show a much wider span of temperature. The higher intensities also allow us to measure temperatures as a function of time during an event.

Figure 7 compares the observed pattern of enhancements early in the event of 22 May 2013 with a plot of *A/Q vs. T*. Unlike the pattern of elements in impulsive events, here C, N, and O have moved well above He to approach Ne, while Mg, Si, and S have moved far above Ne near Ar and Ca. We can almost judge the temperature by inspection from the pattern of abundance enhancements. We could scale He and Fe to match the spread at some other *T*, but the pattern of the intermediate elements would not fit.

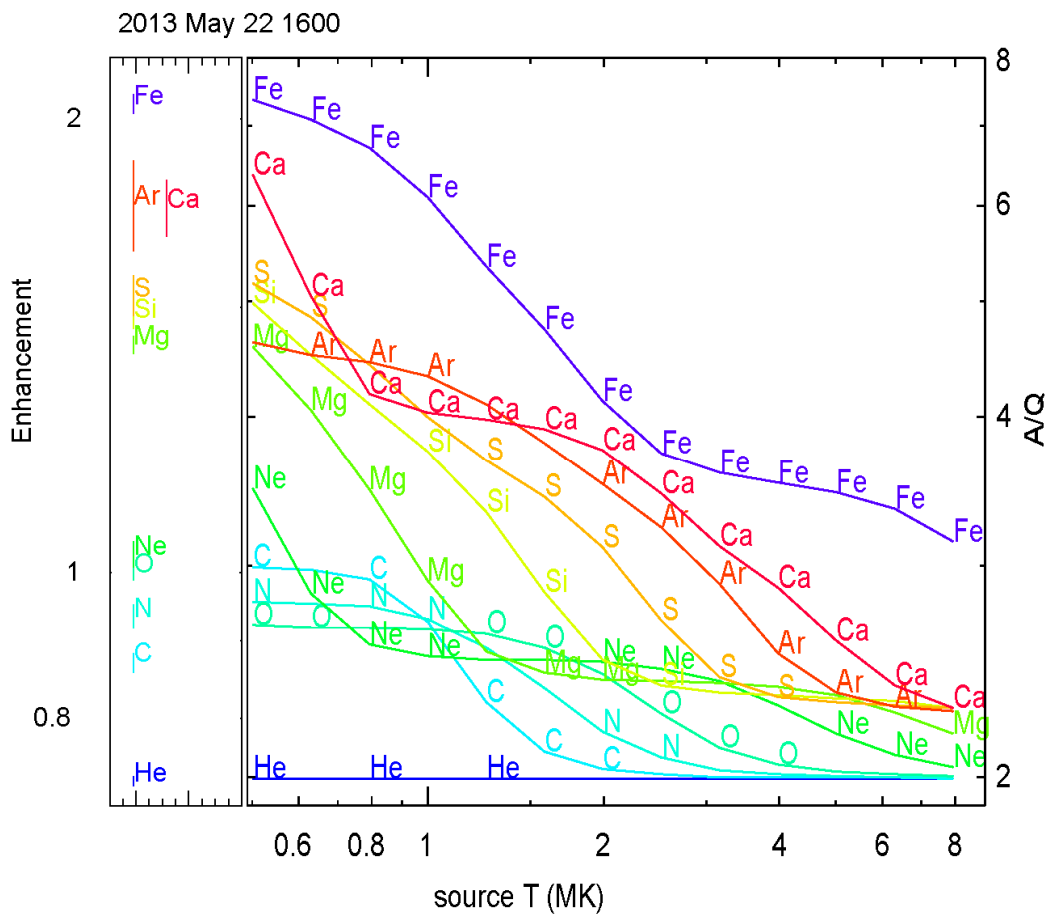

**Fig. 7** The left panel shows observed enhancements in element abundances at *Wind* during the interval 1600 – 2400 UT 22 May 2013. The right panel shows A/Q *vs.* T for various elements, as in Figure 5. The groupings of enhancements match those in *A/Q* near 0.6 – 0.8 MK. Compared with impulsive events, here C, N, and O have moved above He to join Ne while Mg, Si, and S have left Ne approaching Ar and Ca (Reames 2016a).

A more complete analysis of this SEP event of 22 May 2013, in Figure 8, shows the deduced temperature dependence in 8-hr intervals throughout the event in panel (c).

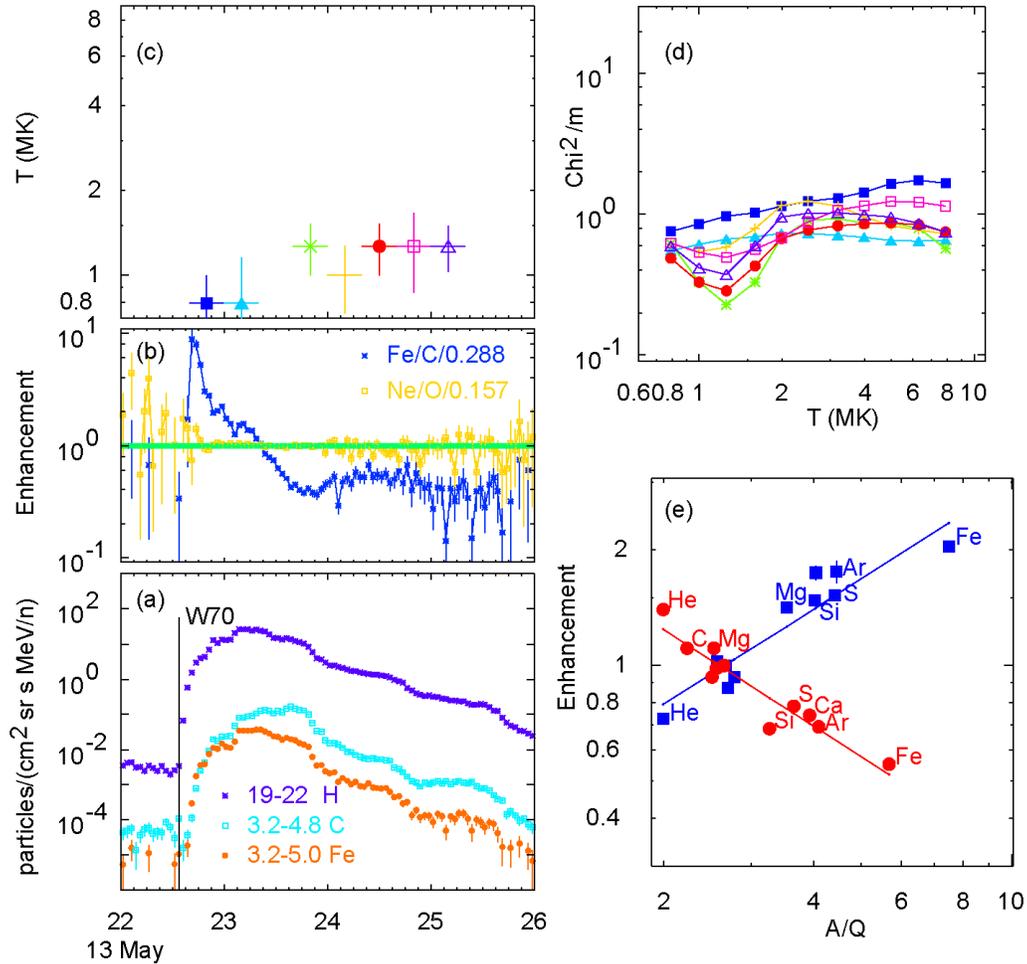

**Fig. 8** Analysis of the 22 May 2013 gradual SEP event shows typical (a) intensities and (b) enhancements *vs.* time, (c) resulting temperatures *vs.* time, color coded, (d) $\chi^2/m$ *vs.* $T$ for each time, and (e) best fits of enhancement *vs.* $A/Q$ at two selected times, early (blue square) and late (red circle). Either enhancement or suppression of abundances may be used to determine *T*, as shown in (e). Temperatures remain fairly constant throughout most events (Reames 2016a).

There are seven time intervals in the event in Figure 8. For each time interval, least-squares fits of enhancement *vs.* $A/Q$ are done, using *each* of the 11 temperatures to determine $A/Q$ values. The resulting $\chi^2/m$ is plotted *vs.* $T$ as in panel (d) of Figure 8; the minimum value of $\chi^2/m$ determines the best *T* for that time.

For example, the fifth time interval, at 0800 – 1600 on 24 May in Figure 8c, is shown as filled red circle. Fits using enhancements during this period and $A/Q$-values at *each* temperature result in the $\chi^2/m$ *vs.* $T$ curve of filled red circles in panel (d). This curve has a minimum at $T = 1.26$ MK, which is then plotted as a filled red circle in panel (c) and the best fit line of enhancement *vs.* $A/Q$ at that temperature is shown along with the points for each element as filled red circles in panel (e).

Note that one time interval has been skipped on 23 May in Figure 8 where the enhancement in Fe/O ≈ 1 in panel (b). When there is neither enhancement nor suppression, the power-law is flat, any $A/Q$ will fit, and we cannot determine a temperature. When abundances are nearly coronal we find uncertain or ambiguous values of *T* – there has not been enough scattering to alter the coronal abundances. Large enhancements or suppressions provide the best temperatures.

In a study of 45 gradual SEP events, 11 events (24%) had temperatures of 2.5 – 3.2 MK just like impulsive events (Reames 2016a). These have been understood as events that are dominated by reacceleration of residual suprathermal ions from impulsive SEP events. Event temperatures have no strong dependence on properties, such as CME speed, as shown in Figure 9, nor would we expect any. Data in the figure show an un-weighted correlation coefficient of -0.49. GLEs are distinguished in the figure but apparently they have no special significance.

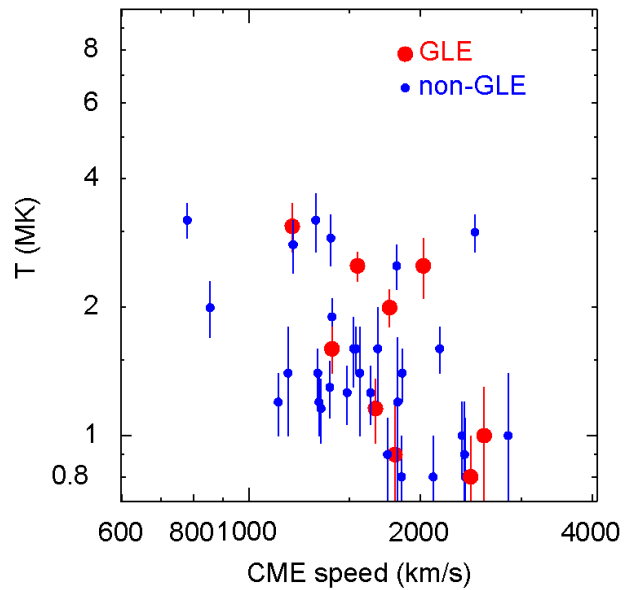

**Fig. 9** Temperatures of the source plasma in gradual SEP events are shown *vs.* speed of the driving CME. GLEs, emphasized in red, are unremarkable (Reames 2016a, 2016c).

### 3.2.3 Comparing Impulsive and Gradual Events

Since we now have temperatures for both impulsive and gradual events, it is possible to compare properties of the two in the same temperature range, especially at high temperatures where He, C, N, and O are all likely to be fully ionized so that their relative abundances cannot be altered by either acceleration or transport.

Figure 10 shows a comparison of enhancements of O/C *vs.* C/He for impulsive (lower panel) and gradual (upper panel) SEP events, with coronal normalization values shown along the axes. The impulsive events are limited only by removal of events with large statistical errors. Gradual events include all 8-hr intervals during events with $2 \leq T \leq 4$ MK.

Note that the gradual events in Figure 10 are tightly distributed while the impulsive events have a large spread with non-thermal errors that exceed the statistical errors shown. If these ~3 MK gradual events are produced by shock reacceleration of suprathermal ions from impulsive SEPs from jets, they must average over the output from many small jets to reduce the non-thermal variance. As we noted previously, strong enhancements in $^3$He/$^4$He are seen even in quiet periods (Desai et al. 2003; Bučík et al. 2014, 2015; Chen et al. 2015). These $^3$He-rich periods must involve many small jets in active regions producing many impulsive SEP events that are too small to be resolved individually. Shock waves passing through these regions would average a seed population from many impulsive events, diluted by some ambient coronal plasma.

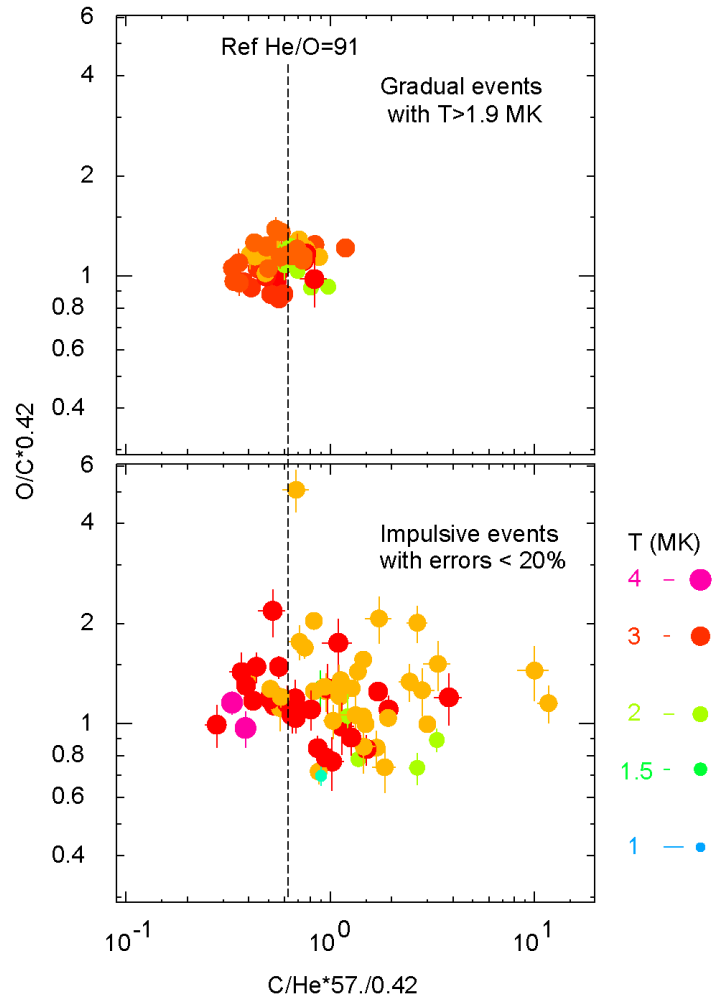

**Fig. 10** Enhancements of O/C *vs.* C/He are compared, for measurements in gradual SEP events with $T > 1.9$ MK (upper panel) and for impulsive SEP events with < 20 % errors (lower panel). *T* is indicated by the size and color of the symbols. Both panels are plotted on the same scale to show differences in the distribution size. The median of the distribution of C/He for the gradual events, shown as a dashed line, implies a reference coronal value for He/O of 91 rather than 57 (Reames 20016b), i.e. He/O = 91 normalized the dashed line to 1.0. The spread in the impulsive events of a single temperature (color) are "non-thermal" variations.

　　Another feature of Figure 10 is the offset of the mean enhancement of C/He for the hot gradual events form one. Since C must be fully ionized at the temperatures in these events, this means we must have the wrong "coronal" or reference value for He. Instead of He/O of 57, as assumed, these hot events would require He/O=91. Note the effect of the higher He/O value as the red circle on the FIP plot in Figure 3.

　　As we move down in temperature for the cooler gradual events, O is no longer fully ionized, and as it is enhanced, He/O decreases. Thus averaging He for all gradual events gives a lower value of He/O. How much do reference abundances vary?

## 4 Reference Abundance Variations

The SEP coronal abundances we have used to measure "enhancements" in both impulsive and gradual events have been derived by averaging over many gradual SEP events, either in solar cycle 21 (Reames 1995) or in cycles 23 and 24 (Reames 2014). This assumes that when differences in particle scattering produce enhancements at one time in one location it will cause depressions at another time and location. Is there a single "coronal" abundance for each element?

## *4.1 Helium Abundance Variations*

Variations of He/O are seen in the solar wind as functions of time and of solar-wind speed (Collier *et al.* 1996; Bochsler 2007; Rakowsky and Laming 2012) and large variations are seen in H/He with phase in the solar cycle (Kasper et al. 2007). Unfortunately, we cannot deduce coronal H from SEPs since H strongly controls its own wave generation and scattering (e.g. Reames, Ng, and Tylka 2000; Ng, Reames, and Tylka 2003), among other problems, while heavier ions are "test particles." Are the solar wind abundance variations a property of the solar corona itself, or only related to the formation of the solar wind? Could these, or other, coronal variations also be measurable in SEPs? Evidence in Figure 10 that the "hot" events could have a reference He/O=91 rather than 57, strongly suggests that this is possible.

In some gradual SEP events there is also evidence that the observed value of the enhancement in He/O does not fit the power-law dependence on *A/Q*. Three examples are shown in Figure 11. For the event in panel (a) He falls below the fits; the fit and $\chi^2$ would be improved using a *reference* He/O ≈ 40. For the event in panel (c), He falls well above the fit lines; the fits and $\chi^2$ would be improved using a reference He/O ≈ 90. The event in panel (b), *and for the event in Figure 8*, reference He/O = 57 seems just right.

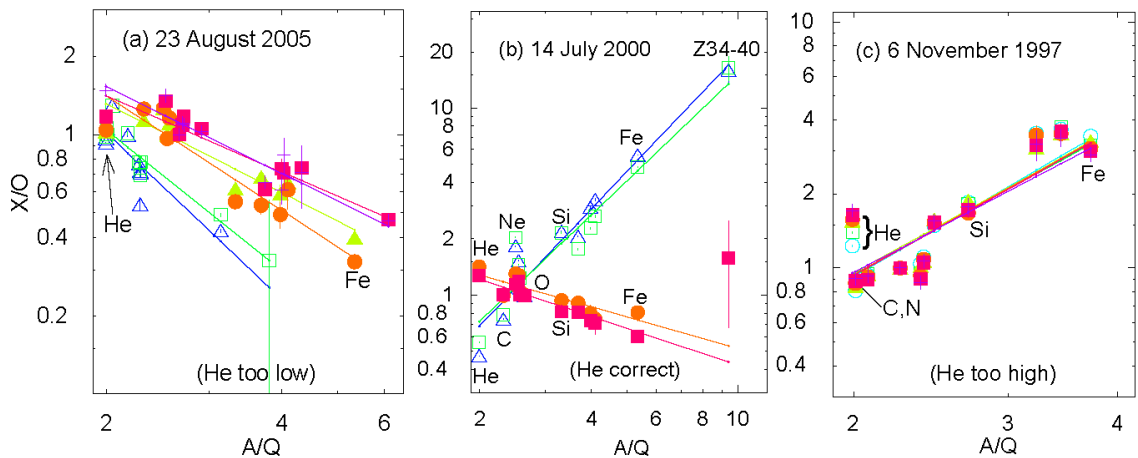

**Fig. 11** Least-squares power-law fits of enhancement *vs. A/*Q are shown during three gradual SEP events. *All* He enhancements shown are based upon a reference value of He/O = 57. In (a) the event of 23 August 2005, He falls below the fit lines; in (b) the event of 14 July 2000, both enhancements and suppressions appear consistent, and in (c) the event of 6 November 1997 He falls well above the fit (based on fits in Reames 2016a).

For each of the 8-hr periods studied in the gradual SEP events, suppose we calculate the value of the reference He/O that would be required to bring the He enhancement to the fit line. Figure 12 is a histogram of the number of occurrences of each reference He/O as a function of *T*. There is clearly a temperature dependence of the *reference* He/O value, in addition to the power-law transport.

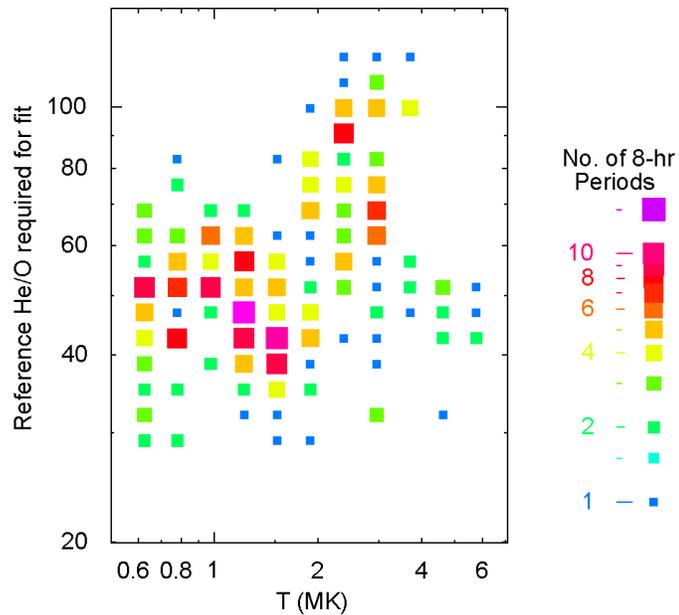

**Fig. 12** A histogram of the reference value of He/O that would be required to bring the He enhancement onto the best-fit power-law is shown *vs. T*. The histogram includes each 8-hr period during gradual events. Temperatures in the 2.5–3.2 MK active-region intervals require higher values of the underlying coronal He/O ratio (Reames 2017c).

The reference value of He/O is quite constant during events and its variation in events throughout the solar cycle is shown in Figure 13. Since the abundances of the hotter gradual events are actually determined from the impulsive suprathermal ions in the seed population, the higher value of He/O may be determined deeper in the corona.

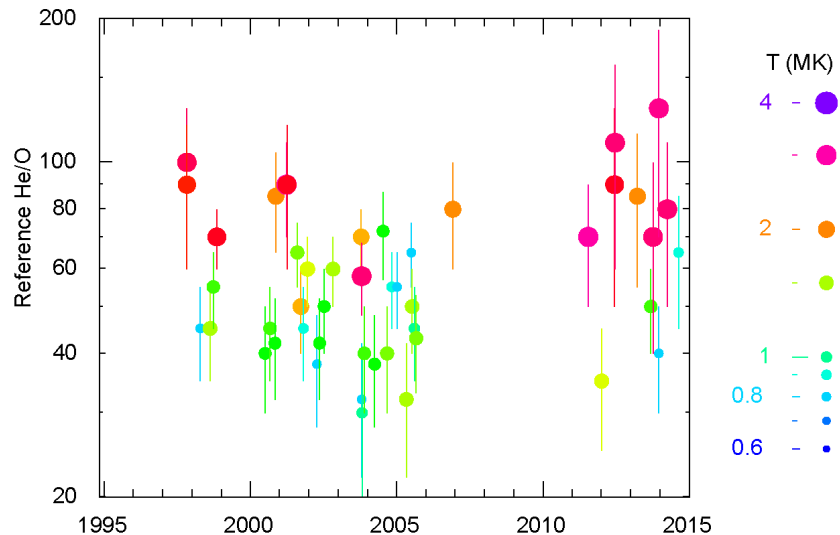

**Fig. 13** The estimated coronal value of He/O required for each gradual SEP event is shown *vs.* time, with the source plasma temperature indicated as the size and color of the circle (Reames 2017c).

### *4.2 Abundances of Heavier Ions*

It is possible to treat the abundances of heavier ions just as we have treated the abundances of He, *i.e.* to ask what reference value would bring the enhancement to the power-law fit line of the other elements. However, other elements show no significant variations. For C/O, for example, mean values show essentially no temperature difference: derived coronal C/O = 0.392±0.007 averaged over SEP events with $T$ = 2–4 MK and C/O = 0.398±0.006 for $T$ < 2 MK (Reames 2017c). However, this ratio for SEPs *never* exceeds 0.50, nor approaches the mean value of C/O = 0.68±0.07 in the solar wind (Bochsler 2009). C/O shows the largest significant difference between SEPs and the slow solar wind, a difference that has been unexplained, although a more recent measurement in the bulk solar wind does give C/O=0.53±0.06 (Heber *et al.* 2013). Why do SEPs and the slow solar wind disagree?

## 5 Coronal Abundances: SEPs and the Solar Wind

An improved understanding of the He abundance in SEPs has allowed a new analysis of measurements of coronal abundances of the elements. The lower panel in Figure 14 compares the average SEP element abundances (e.g. Reames, 2014, 2017a, 2018) with those of the slow or interstream solar wind (SW; Bochsler 2009) by showing the ratio of the two as a function of FIP. SEP abundances are those shown in Figure 3 and listed in Table 1, with He/O = 91. The upper panel shows both SEP and SW abundances relative to the photospheric abundances with the dominant element species from Caffau et al. (2011) and other elements from Lodders, Palme, and Gail (2009) just as in Figure 3. Data shown in Figure 14 are listed in Table 1.

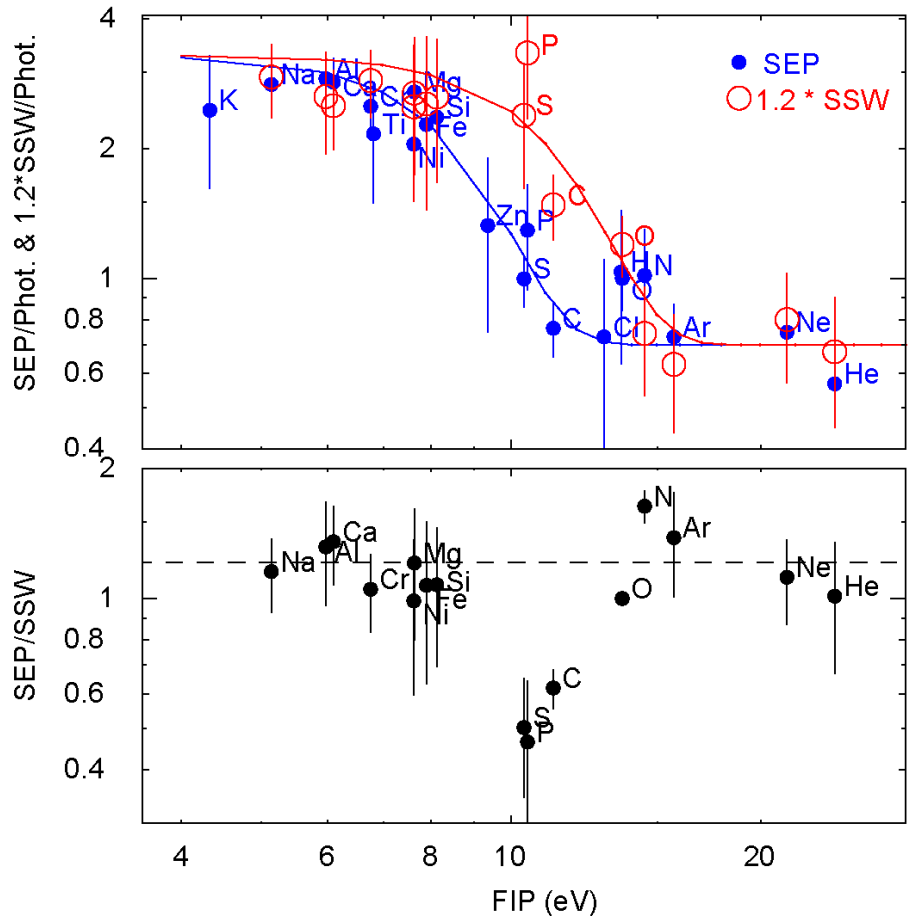

**Fig. 14** The upper panel shows the SEP/photospheric and 1.2 × slow solar wind (SSW)/photospheric abundance ratios as a function of FIP. Curves are empirical curves used to help show the trend of each data set. The lower panel shows the ratio of the "coronal" abundances from SEPs to those of the slow SW (Bochsler 2009), as a function of FIP. The dashed line suggests the preferred normalization factor of 1.2. Data shown in Figure 14 are listed in Table 1 (Reames 2018).

Table 1 Photospheric, SEP, CIR, SSW, and FSW Abundances.

| | Z | FIP [eV] | Photosphere[1] | SEPs[2] | CIRs[3] | Interstream Solar Wind[4] | Coronal Hole Solar Wind[4] |
|---|---|---|---|---|---|---|---|
| H  | 1  | 13.6 | 1.74×10⁶ * | (≈1.6±0.2)×10⁶ | (1.81±0.24)×10⁶ | – | – |
| He | 2  | 24.6 | 1.6×10⁵ | 91000±5000 | 159000±10000 | 90000±30000 | 75000±20000 |
| C  | 6  | 11.3 | 550±76* | 420±10 | 890±36 | 680±70 | 680±70 |
| N  | 7  | 14.5 | 126±35* | 128±8 | 140±14 | 78±5 | 114±21 |
| O  | 8  | 13.6 | 1000±161* | 1000±10 | 1000±37 | 1000 | 1000 |
| Ne | 10 | 21.6 | 210 | 157±10 | 170±16 | 140±30 | 140±30 |
| Na | 11 | 5.1 | 3.68 | 10.4±1.1 | – | 9.0±1.5 | 5.1±1.4 |
| Mg | 12 | 7.6 | 65.6 | 178±4 | 140±14 | 147±50 | 106±50 |
| Al | 13 | 6.0 | 5.39 | 15.7±1.6 | – | 11.9±3 | 8.1±0.4 |
| Si | 14 | 8.2 | 63.7 | 151±4 | 100±12 | 140±50 | 101±40 |
| P  | 15 | 10.5 | 0.529±0.046* | 0.65±0.17 | – | 1.4±0.4 | – |
| S  | 16 | 10.4 | 25.1±2.9* | 25±2 | 50±8 | 50±15 | 50±15 |
| Cl | 17 | 13.0 | 0.329 | 0.24±0.1 | – | – | – |
| Ar | 18 | 15.8 | 5.9 | 4.3±0.4 | – | 3.1±0.8 | 3.1±0.4 |
| K  | 19 | 4.3 | 0.224±0.046* | 0.55±0.15 | – | – | – |
| Ca | 20 | 6.1 | 3.85 | 11±1 | – | 8.1±1.5 | 5.3±1.0 |
| Ti | 22 | 6.8 | 0.157 | 0.34±0.1 | – | – | – |
| Cr | 24 | 6.8 | 0.834 | 2.1±0.3 | – | 2.0±0.3 | 1.5±0.3 |
| Fe | 26 | 7.9 | 57.6±8.0* | 131±6 | 97±11 | 122±50 | 88±50 |
| Ni | 28 | 7.6 | 3.12 | 6.4±0.6 | – | 6.5±2.5 | – |
| Zn | 30 | 9.4 | 0.083 | 0.11±0.04 | – | – | – |

[1] Lodders, Palme, and Gail (2009).
* Caffau *et al.* (2011).
[2] Reames (1995, 2014, 2017a, 2018).
[3] Reames, Richardson, and Barbier (1991); Reames (1995).
[4] Bochsler (2009).

Compared with the SSW, SEP abundances of P, S, and C are low, and N is high; all other element abundances are consistent as shown in Figure 14. Apart from a small vertical shift based upon differences in the O normalization, the greatest difference between the SEP and SW abundances is a difference in the FIP of the transition from low- to high-FIP abundance levels. One could argue that the transition from low- to high-FIP levels changes from ~10 eV for SEPs to ~14 eV for the SSW, so that P and S, and even C, tend to be high-FIP elements for SEPs but clearly low-FIP elements for the SW. The need for the factor of 1.2 implies that even O is elevated somewhat by the higher crossover for the SW. The difference in crossover seems to require a completely different region of FIP processing for the two populations as noticed previously by Mewaldt et al. (2002).  Naively, we might suggest that the photosphere is cooler, on average, in sunspots beneath active regions producing SEPs, so that C, S, and P are neutral atoms there, while the SW comes from warmer photospheric plasma, where S and P, at least, are ionized, and C is partially ionized.  However, the differences are already explained well by current theory, where ions are much more easily swept up into the corona than neutral atoms as by the ponderomotive force of Alfvén waves (Laming 2004, 2009, 2015).

The ions that will eventually be shock-accelerated to become SEPs originate on closed magnetic loops where Alfvén waves can resonate with the loop length and the fractionation is concentrated near the top of the chromosphere where H is becoming ionized (see Figure 8 and Table 3 of Laming 2015), restricting fractionation, especially of C, S, and P.  Ions that will be SSW begin on open field lines where waves produce ponderomotive force further down where H is neutral and fractionation is easier, particularly for C, P, and S.  For the SSW, the amplitude of the FIP-bias depends upon the amplitude of slow-mode acoustic waves as shown in Table 4 of Laming (2015).

Theory and observations were compared by Reames (2018) as shown in Figure 15.  The lower panel compares the SEPs with the theory of closed field lines (Table 3 of Laming 2015), while the upper panel compares the SSW and energetic ions from corotating interaction regions (CIRs) with the theory for open field lines (Table 4 of Laming 2015).  CIRs occur where high speed solar-wind streams overtake low-speed wind emitted earlier in the solar rotation (e.g. Richardson 2004).  Two shock waves are formed at the CIR, a forward shock propagates outward into the slow wind and reverse shock propagates sunward into the fast wind.  Energetic particles are mainly accelerated at the stronger reverse shock.  The upper panel in Figure 15 shows that C and S are enhanced in the accelerated CIR ions just as they are in the SSW.  For the SEPs, C/O seems somewhat more suppressed than it is in the theory.

The origins of coronal SEP and SW ions have another important difference.  The corona is weakly bound to the Sun by a collection of high closed magnetic loops that help to contain the plasma.  A fast CME-driven shock wave driving through the corona can rapidly accelerate some of the local ions to MeV-energy SEPs.  The *local magnetic fields are no longer strong enough to contain these new SEPs, so they easily escape*, beginning at about $2 - 3$ $R_S$ (Reames 2009b).  The *SW ions cannot sample these closed loops* unless they are eventually opened; SW ions must originate on field lines that are truly open to 1 AU and beyond.  In addition, high loops tied more closely to solar jets in active regions may help to trap impulsive suprathermal ions for later shock acceleration.  However, in the Laming (2015) theory, the FIP pattern is determined by closed or open fields at the base of the corona; it is irrelevant whether the fields remain closed or open later.

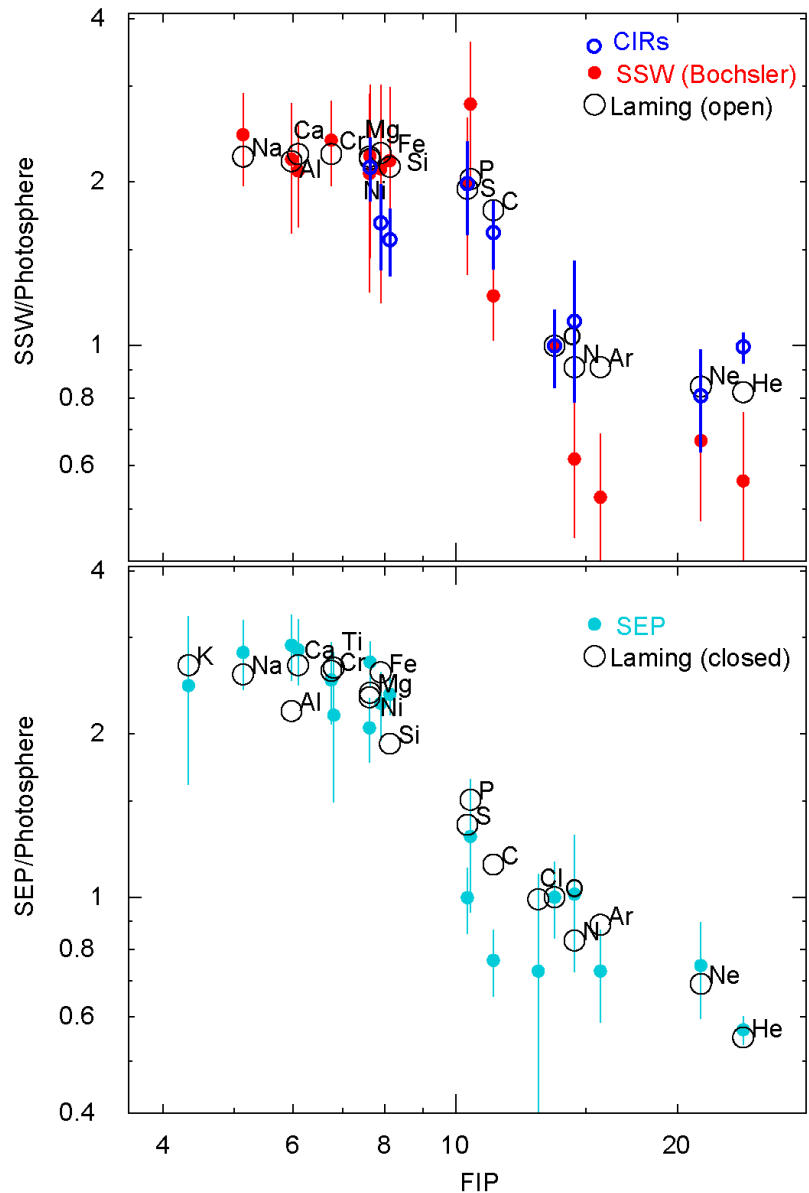

**Fig. 15**. The lower panel compares the FIP pattern of SEPs with the closed loop model of Laming (2015), Table 3. The upper panel compares the SSW and CIR FIP patterns the open field model of Laming (2015), Table 4 (Reames 2018)

Recently, extreme-ultraviolet (EUV) and X-ray spectroscopy observations near sunspots have shown regions with variations in the abundance ratios of Ar and Ca that sometimes even involve a reverse-FIP effect where Ar/Ca is enhanced (Doschek and Warren 2016, 2017). While this may also suggest the effects of a cooler environment, it is different in character from the SEP-SW differences above. In our case the amplitude of the FIP level difference varies little from SEPs to the SW; Ca/Ar = 2.6 ± 0.4 in SEPs and 2.6 ± 0.7 in the SW. These elements are not in the region of crossover from high to low FIP. The difference between the SEP and SW FIP patterns is almost entirely in the value of the crossover FIP.

It is somewhat surprising that we never see evidence of SEPs with FIP-bias like that of the SW, since shocks *do* propagate even into the fast solar wind, (Kahler and Reames 2003, Kahler, Tylka, and Reames 2009). Perhaps the density of seed particles trapped in filaments and near active regions, coupled with cross-field diffusion in the shock front, swamps direct acceleration from the SW.

Brooks, Ugarte-Urra, and Warren (2016) have identified regions of high FIP-level abundance differences on the Sun using spectral line measurements to determine Si/S abundance ratios which they show for the whole Sun. The Si/S ratio tends to be enhanced in the vicinity of active regions. While these measurements do indeed determine regions of interest for SEP acceleration, they have much less relevance for the slow solar wind where Si and S are both low-FIP ions and their radio does not measure FIP level differences. Si/S may show source regions for SEPs, but does not show the source region of the slow solar wind. A different choice of spectral lines, e.g. for Mg/Ne, if possible, might be a better indicator of the origin of both SEPs and the solar wind when compared with Si/S.

We consider the sketch of a possible magnetic configuration to produce the observed SEP-SW FIP differences in the Discussion section below.

## 6 Spatial Distributions – Multi-Spacecraft Observations

Now that we can use element abundances to measure source-plasma temperatures, we can study particle source properties with multiple spacecraft and address new questions. Do the SEPs at each longitude have a local origin? To what extent do particles from a common origin spread laterally across the face of a shock wave? We can compare SEP abundances observed over a broad longitude span at the *Wind* and the two STEREO spacecraft. Some basic properties of the events at STEREO, such as Fe/O spectral variations, peak intensities, and fluences have been studied recently by Cohen et al. (2014) and Cohen, Mason, and Mewaldt (2017).

Figure 16 shows an analysis of the interesting event of 31 August 2012 at *Wind* and STEREO B. The properties of the event at both spacecraft are seen in panels (a) and (b) while the resulting temperatures are seen in panel (c). The $\chi^2$ curves for STEREO B in panel (d) show an early minimum at high *T* then transition to lower *T*, while those for *Wind* in panel (e) always minimize at low *T* throughout the event.

Early in the event, STEREO B, which is magnetically well-connected to the source, sees plasma at 3.2±0.8 MK while *Wind*, 116° to the west, sees 1.6±0.2 MK, as seen in Figure 16c. Figure 16f shows the change in the power-law fit at STEREO B with time. When the temperature at STEREO B declines on 3 September (panel (c)), all of the intensities (panel (a)) on both spacecraft are quite similar, suggesting that they are both in a reservoir region (e.g. Reames 2013, 2017a) behind the CME. Reservoirs function as a magnetic bottle behind a CME, where spatially uniform intensities decrease adiabatically with time as the CME expands, increasing the volume of the bottle. These observations suggest that the source of the plasma in the reservoir differs from that seen early in the event, i.e. the change in time is caused by a variation in space.

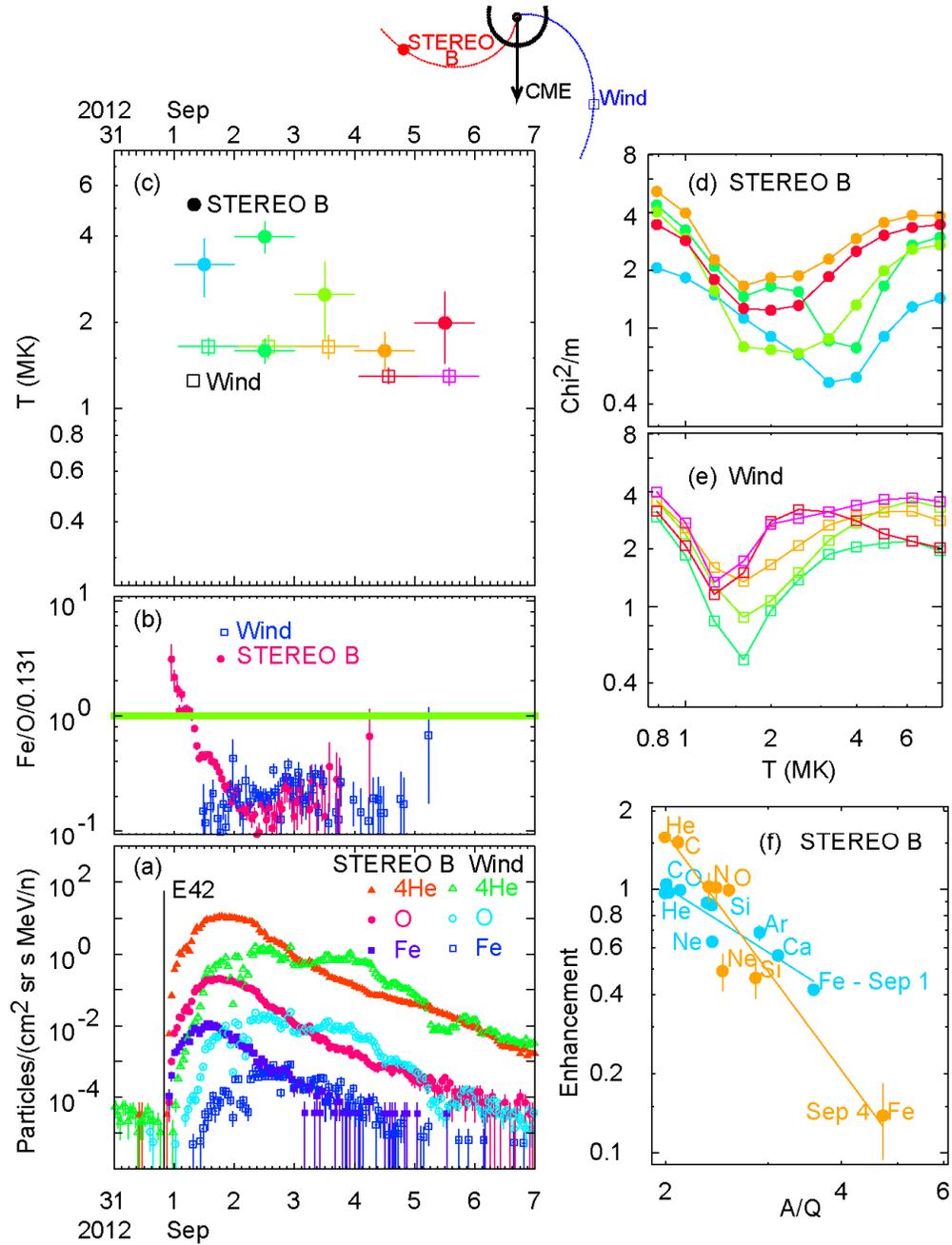

**Fig. 16** The 31 August 2012 SEP event is shown for STEREO-B (filled) and *Wind* (open symbols): (a) the intensities of He, O, and Fe *vs.* time, (b) the enhancements of Fe/O *vs.* time, (c) the derived source temperatures, *T vs.* time, (d and e) $\chi^2/m$ *vs. T*, color coded for time, for (d) STEREO-B and (e) *Wind*, and, (f) enhancement *vs. A/Q* fits for early and late days at STEREO-B. The spacecraft locations relative to the CME are shown above (Reames 2017b)

　　　It is possible to use the two parameters, slope and intercept, of each fit to map enhancements of the elements into *A/Q*-space to see how the individual element abundances compare with expected *A/Q* values at the best-fit temperature. Fig. 17 shows such a comparison on 1 September 2012 for enhancements at *Wind* and STEREO B.

**Fig. 17** Measured values of enhancements, X/O at *Wind* and STEREO-B on 1 September 2012, are mapped into the theoretical plot *A/Q vs. T* at 1.6 MK and 3.2 MK, respectively, using the best-fit parameters for each spacecraft. These measurements correspond to different source plasma temperatures at different locations in space, as shown in the inset, at the same time early in the SEP event (Reames 2017b).

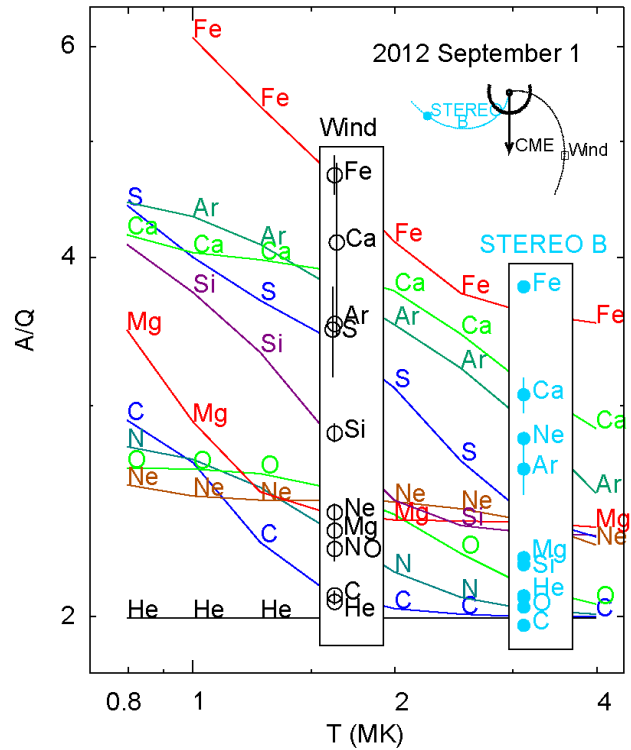

The abundance enhancements at *Wind* match the pattern of *A/Q* quite well: He and C are fully ionized but N and O have joined Ne and Mg while Si and S have moved up. STEREO B, shows the impulsive pattern with He, C, N, and O all fully ionized (in fact, He should have been divided by 91 rather than 57 so it is not above C and O); however, Ne is overly enhanced relative to Mg and Si. Actually, of course, these are suppressions that decrease with increasing *A/Q*, not enhancements (e.g. see Figure 16f) and there are other events with non-thermal variations in elements other than Ne. However, late in the event, the same anomaly in Ne/O is seen at both *Wind* and STEREO B, again suggesting a reservoir.

So far we only have one event with clear spatial variations in source temperatures. Three other events that were studied show possible variations, but temperature differences were not clearly resolved statistically (Reames 2017b).

# 7 Discussion

Our current understanding of impulsive SEP events is as follows: A steep power-law dependence of the average element abundance enhancements on *A/Q* is seen to be similar at low energy as $(A/Q)^{3.3}$ (Mason et al. 2004) and higher energies as $(A/Q)^{3.6}$ (Reames, Cliver, and Kahler 2014a), across the periodic table. This power law appears to be an expected consequence of scattering of ions in islands of magnetic reconnection (as found in particle-in-cell calculations by Drake et al. 2009) that occur on open magnetic field lines in solar jets (Kahler, Reames, and Sheeley 2001) accompanied by slow, narrow CMEs without shocks (Reames, Cliver, and Kahler 2014a). Copious electron acceleration in the reconnection region leads to electron beams that produce type III radio bursts that have long been associated with these impulsive events that were first identified as $^3$He-rich events (Reames, von Rosenvinge, and Lin 1985; Reames and Stone 1986). The huge enhancements of $^3$He appear to be produced by resonant wave-particle interactions with electromagnetic ion-cyclotron waves that may be produced by the streaming electrons (Temerin and Roth 1992).

The electron temperature in the plasma that is accelerated in impulsive SEP events is ≈ 3 MK as determined from the pattern of ionization states $Q$ consistent with the power law dependence upon $A/Q$ (Reames, Cliver, and Kahler 2014a, 2014b). However, the reconnection region may occur below 1.5 solar radii, so that the ions traverse enough material after acceleration to strip them to equilibrium charge states that depend upon their energy (DiFabio et al. 2008), but the amount of material traversed is not enough to cause energy loss of the extremely high-Z ions, since that energy loss would have destroyed the power-law dependence on $A/Q$.

Abundance measurements in SEP events also include isotope abundances of the elements from C through Ni that have been reviewed by Leske et al. (2007). The abundance ratio $^{22}$Ne/$^{20}$Ne, for example, in many large gradual events is similar to that found in the solar wind, but it tends to follow the behavior of Fe/O, increasing for events with increased Fe/O. Comparing neighboring elements, $^{22}$Ne/$^{20}$Ne is correlated with Na/Mg. $^{22}$Ne/$^{20}$Ne is also correlated with $^{26}$Mg/$^{24}$Mg, with $^{56}$Fe/$^{54}$Fe, and with most other isotope ratios. The isotope variations seem to be an extension of the general $A/Q$ variations, following differences in $A$, rather than $Q$. All these $A/Q$ enhancements are much larger in impulsive SEP events. Most isotopic abundances are within ~10% of solar-system values with the possible exception of $^{13}$C/$^{12}$C which may be enhanced by a factor ~2 (Leske et al. 2007).

The average abundances, the power-law enhancements, and source plasma temperatures seem well defined for most events, both above and below 1 MeV amu$^{-1}$. However, there are a few small impulsive events with steep spectra that have unusual abundances, such as an unusual enhancement of N and especially of S (Mason, Mazur, and Dwyer 2002; Mason et al 2016), in one event S/O = 32. Rounded energy spectra of some of the heavy nuclei in these events, near 0.1 MeV amu$^{-1}$, are similar to the rounded spectra of $^3$He rather than the power-law energy spectra of $^4$He or O (Mason et al 2016). These events are rare, about one per year, and most are too small to be measureable above 1 MeV amu$^{-1}$. However, we do not understand these events, nor do we understand the source of the non-thermal abundance variations in impulsive events (e.g. Figure 10).

It is commonly known that impulsive SEP events occur in clusters (e.g. Reames, von Rosenvinge, and Lin 1985; Reames and Stone 1986; Reames 2000; Reames 2017a) while an observer is magnetically connected to a single active region on the Sun. Desai et al. (2003) found $^3$He-rich periods that were otherwise quiet that lasted for days. Multi-day and recurrent $^3$He-rich periods have been associated with active regions (Bučík et al. 2014, 2015; Chen et al. 2015). These periods are probably caused by many small jets each producing a $^3$He-rich SEP event that is too small to be resolved as an individual event. As we move toward smaller events, they become more numerous. A similar situation occurs with flares, where flare energy decreases, the number of flares per day increases as a power law. This observation led Parker (1988) to suggest that nanoflares were sufficient to heat the corona. Flares involve magnetic reconnection on closed field lines while jets are from reconnections on open field lines, so it seems plausible that "nanojets" might keep a basic level of $^3$He-rich material flowing from active regions.

When the fast shock wave of a gradual SEP event passes above these active regions, it encounters a seed population that has averaged suprathermal ions over many jets so the non-thermal variations in abundance are small, as seen in Figure 10 (Reames 2016b). In Figure 10, temperature is indicated by color, so the large abundance variations in impulsive events of a single color, in the lower panel, must be non-thermal. Averaging over just 10 impulsive events can reduce a 30% variation to 10%.

Figure 18 is a sketch of a possible configuration leading to different FIP dependence of SEP and SW abundances. Field lines from the active-region corona (red) connect to closed loops or once-closed loops where chromospheric C atoms are highly neutral and S and P are partially so. Red field lines carry the SEP pattern of FIP. Recurrent solar jets often contribute impulsive suprathermal ions to this region. Field lines carrying slow SW (yellow, green) originate outside the active region where open field lines from the photosphere outside active regions fractionate S and P and some of the C to produce the FIP pattern of the slow SW (Figures 14 and 15). In the upper panel of Figure 18, a CME-driven shock wave accelerates material from the high corona above an active region, possibly including residual impulsive suprathermal ions from previous jets (enhanced in $^3$He and heavy ions), left-over bulk ejecta from jets (unenhanced), and other material from the high corona above the active region. Some regions of the shock wave in many gradual SEP events must sample only recently-formed high coronal loops, where the slowly-evolving He abundance has attained as little as half of its full value (Rakowski and Laming 2012; Laming 2015). As the CME and shock expand, both radially and laterally, with time they push aside the neighboring regions and even deflect streamers (Rouillard et al. 2011, 2012) spreading the SEPs over a wider longitude range.

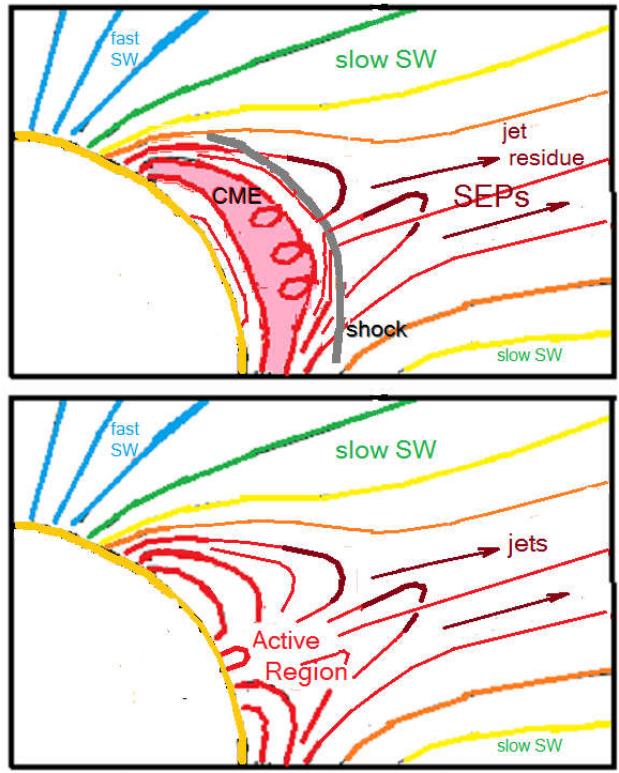

**Fig. 18** Sketch of possible sources of SEPs and the SW. The lower panel shows an active region (red), containing closed loops, from which solar jets emerge with a SEP-like FIP pattern. Field lines carrying the slow SW (yellow and green) diverge from open field lines from the photosphere outside of active regions. The upper panel depicts a CME-driven shock wave (gray) that accelerates material from the high corona and residue from jets. Blue field lines track the fast SW from coronal holes that have a photospheric source similar to the slow SW, but less trapping and divergence and a weaker FIP effect. Newly-formed He-poor closed coronal regions on the fringe of active regions are not shown.

The average element abundances in gradual SEP events are a measure of element abundances in the closed-field active-region solar corona. The acceleration by shock waves in gradual SEP events begins at an altitude of 2 – 3 solar radii (Reames 2009a, 2009b). These abundances are comparable with those from spectral line measurements of corona and with those of the slow solar wind (Schmelz, 2012). However, the origin of the differences between the SEP and SW abundance is *not* in the corona, where the SEP ions are accelerated; it must lie in the chromosphere that underlies the active regions where most of the SEP acceleration takes place, altering the location of the crossover on a FIP plot as in Figure 14. Resonance of Alfvén waves with the loop length of closed field lines modifies the altitude where their ponderomotive force fractionates the ions.

Gradual SEP events involve shock acceleration of the ions from the ambient solar corona at ~1 – 2 MK or the acceleration of remnant suprathermal ions at ~3 MK from impulsive SEP events above active regions as just described. The use of the *A/Q*-dependence of abundance enhancements works well in the energy region of a few MeV amu$^{-1}$. At higher energy, usually above ~10 MeV amu$^{-1}$, the spectra soon break downward and steepen, introducing abundance variations that depend upon the seed population and the shock geometry (e.g. Tylka et al. 2005; Tylka and Lee 2006). Below 1 MeV amu$^{-1}$ the spectra and abundances depend more strongly upon transport in large events and the shocks sample material far out into the solar wind, but low energies have been ideal for the study of impulsive SEP events (Mason 2007).

In one case, measurement at different longitudes in a gradual SEP event showed 3.2-MK source plasma at one well-connected spacecraft but 1.6-MK source plasma at another, 116° away. In a reservoir region behind the CME temperatures approach 1.6 MK at both spacecraft (Figure 16) and a similar pattern of abundances. Other events show possible, but less definite, temperature variations along the shock. A larger sample of events would be helpful.

Finally, we should note that using abundances to measure temperatures is not a substitute for other direct measurements. The SEP sources are not isothermal, yet we are only able to derive a single temperature. Direct measurements give us important information on the *distribution* of charge states which is also important for understanding the physics. The transport of $Fe^{+10}$ is much different from that of $Fe^{+20}$; in such a mixture, the $Fe^{+10}$ would forge ahead, leading to a substantial increase in $Q_{Fe}$ and in the apparent value of T with time. If we could follow it, the time dependence of $Fe^{+10}/Fe^{+20}$ would behave much like Fe/O that we observe today. In some cases, perhaps, the differences within an event could be as large as those between events of greatly different temperatures. We need to know this, and direct measures of *Q* are still very important.

## 8 Summary

Early observers began the study of SEPs with the largest events, from ground-level where no SEP abundance measurements are possible. Measurements in space soon began to distinguish the unusual abundances in the $^3$He-rich events and the FIP-related coronal-abundance basis of the gradual events. Abundance enhancements that vary as *A/Q* in both event types have led to the use of abundances to measure ionization states and temperatures. The following properties of SEPs are derived from abundances:

1) Lack of detectible abundances of $^{2}$H, $^{3}$H, Li, Be, and B show that SEPs must not have escaped from the nuclear reactions that are seen to occur in solar flares.

2) Average element abundances in *gradual* SEP events provide a measure of coronal abundances.

3) Relative to coronal abundances, enhancement in element abundances in *impulsive* SEP events have a strong power-law dependence on *A/Q*, varying 1000-fold from He to Au or Pb, produced during acceleration in islands of magnetic reconnection in solar jets.

4) 1000-fold increases in $^{3}$He/$^{4}$He are produced in *impulsive* SEP events by resonant wave-particle interactions in the jet sources.

5) Power-law dependence of abundance enhancements on *A/Q*, produced by either acceleration or transport, can be used to estimate *Q*-values and source plasma temperatures. This technique may be used wherever abundance enhancements or suppressions are observed, including remote spacecraft.

6) Residual suprathermal impulsive SEP ions flowing from multiple jets in active regions can be reaccelerated along with ambient plasma by a fast shock wave to produce impulsive abundances with reduced variations.

7) Compared with abundances in the slow solar wind (Bochsler 2009) the largest statistically important discrepancy is C/O = 0.68±0.07 in the solar wind and C/O = 0.42±0.01 in SEPs. This difference, and differences in the abundances of P and S, can be largely explained by differences in the ponderomotive forces of Alfvén waves in open (SW) and closed (SEPs) field lines in the chromosphere. These differences make C, P, and S behave as neutral atoms in SEP sources, but as ions in the solar wind.

8) Some regions of the shock waves in many gradual SEP events also sample recently-formed high coronal loops where the slowly-evolving He abundance has attained as little as half of its full value.

9) Maps of Si/S abundance in the corona show enhancements near active regions that are the likely origin of SEPs, but Si/S enhancements cannot show the origin of the slow solar wind since both Si and S are low-FIP ions in the solar wind.

Abundances of elements and isotopes tell a rich story of the physical processes that characterize SEPs and their sources.

The author thanks Ed Cliver, Steve Kahler, and Martin Laming for helpful discussions and comments on this manuscript.